\documentclass[aps,prb,twocolumn,groupedaddress]{revtex4-1}
\usepackage{graphicx}
\usepackage{dcolumn}
\usepackage{bm}
\usepackage{amsmath}
\usepackage{amssymb}
\usepackage{color}

\begin{document}

\title{Spatiotemporal Patterns in Active Four-State Potts Models}

\author{Hiroshi Noguchi}
\email[]{noguchi@issp.u-tokyo.ac.jp}
\affiliation{Institute for Solid State Physics, University of Tokyo, Kashiwa, Chiba 277-8581, Japan}


\begin{abstract}
Many types of spatiotemporal patterns have been observed under nonequilibrium conditions.
Cycling through four or more states can provide specific dynamics, such as the spatial coexistence of multiple phases.
However, transient dynamics have only been studied by previous theoretical models, since absorbing transition into a uniform phase covered by a single state occurs in the long-time limit.
Here, we reported steady long-term dynamics using cyclic Potts models, wherein nucleation and growth play essential roles.
Under the cyclic symmetry of the four states, the cyclic changes in the dominant phases
and the spatial coexistence of the four phases are obtained at low and high flipping energies, respectively.
Under asymmetric conditions, the spatial coexistence of two diagonal phases appears in addition to non-cyclic one-phase modes.
The circular domains of the diagonal state are formed by the nucleation of other states, and they slowly shrink to reduce the domain boundary.
When three-state cycling is added, competition between the two cycling modes changes the spatiotemporal patterns.
\end{abstract}

\maketitle

\section*{Introduction}

Spatiotemporal patterns, such as wave propagation and spatial chaos, have been observed in various nonequilibrium systems\cite{mikh94,mikh96,rabi00,murr03,kura84,mikh06,beta17,kond21,bail22,meri21,nogu24c,bode00,vana09}.
Characteristic patterns, such as target and spiral waves, have been reproduced theoretically using deterministic continuum equations.
Thermal noise plays an important role in small systems; however, its effects have not been investigated so far.
Experimentally, noise can be added by a random variation in light intensity in the photosensitive Belousov--Zhabotinsky reaction,
and the wave propagation in a subexcitable condition can be enhanced by the noise\cite{mikh06,kada98,alon01}. 

Predator--prey systems exhibit various spatiotemporal patterns and have been extensively studied.
Lattice-based predator--prey models (also called lattice Lotka--Volterra models) consider noise as a random selection of action pairs.
Three species with the rock--paper--scissors relationship can form spiral waves\cite{szol14,szab02,reic07,szcz13,kels15}.
A system involving four or more species can exhibit complicated dynamics, such as the long-lived spatial coexistence of two phases\cite{szol14,dobr18,szab04,szab08,roma12,rulq14,baze19,zhon22,yang23,szol23}.
However, only one of the species survives in the long term, because it multiplies by self-production in the predator--prey systems;
if one species becomes extinct, it is never born again.
Hence, absorbing transition into a uniform phase covered by a single species occurs eventually.
Thus, the short-term dynamics have been studied and the extinction rates have been discussed.

This study employs the active Potts model to clarify the long-term steady dynamics of four-state cyclic systems.
This model involves thermal fluctuations and can be tuned from the thermal-equilibrium (i.e., the original Potts model\cite{pott52,bind80}) to far-from-equilibrium conditions.
We previously studied a three-state cyclic Potts model\cite{nogu24a,nogu24b}.
In a square lattice, each site takes three states ($s=0$, $1$, and $2$) and flips via a Monte Carlo (MC) procedure.
Sites in the same state are attractive and lead to phase separation.
The flipping energies follow the rock--paper--scissors relationship, and one cycle consumes an energy, which can be considered as
the bulk reaction energy in reactions on a catalytic surface\cite{ertl08,bar94,goro94,barr20,zein22} and chemical potential difference for molecular transport through a membrane\cite{tabe03,miel20,holl21,nogu23}.
For the cyclically symmetric condition\cite{nogu24a},
the system exhibits homogeneous-cycling (HC) and spiral-wave (SW) modes at low and high flipping energies, respectively.
In the HC mode, the lattice is dominantly covered by one of the states most of the time,
and the dominant state undergoes cyclic changes ($s=0\to 1 \to 2\to 0$).
In large systems, the HC mode changes discontinuously into the SW mode with increasing flipping energy,
while the two modes coexist temporally in the intermediate flipping energies in small systems and/or at low contact energies.

An non-cyclic one-state phase is obtained through the temporal coexistence of it and SW mode with increasing asymmetry of the flipping energies\cite{nogu24b}. 
The amoeba-like locomotion of biphasic domains is also obtained.
An increase in the flipping energy between two successive states (e.g., $s=0$ and $1$), 
while keeping the other energies constant, induces the formation of the third phase ($s=2$).
Under asymmetric contact energies with symmetric flipping energies, non-cyclic one-state phases and hysteresis are obtained\cite{nogu24b}.
In contrast, in the four-state cyclic Potts model, different modes appear because of the lack of direct interaction between diagonal states.

\section*{Results}

\subsection*{Theoretical Analysis for Homogeneous Mixed States}

\begin{figure*}
\includegraphics{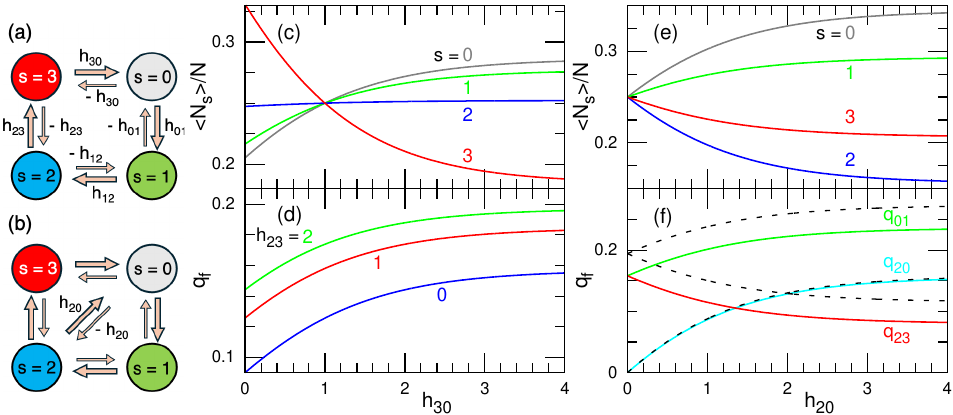}
\caption{
Site flipping in four states ($s=0$, $1$, $2$, and $3$).
(a--b) Schematic of the two models.
(a) Cyclic flips in four states  ($s=0\to 1 \to 2 \to 3 \to 0$).
(b) Addition of three-state cycling ($s=0\to 1 \to 2 \to 0$).
A diagonal flip between $s=0$ and $2$ is allowed.
The wide and narrow arrows represent the forward (dominant) and backward (secondary) flips, respectively.
(c--f) Steady states in the absence of interactions between neighboring sites
under (c--d) the four-state cycling
and (e--f) three- and four-state cycling condition.
(c) Densities of four states as a function of $h_{30}$ at $h_{01}=h_{12}=h_{23}=1$,
given by Eq.~(\ref{eq:st4}).
(d) Flow rate $q_{\mathrm {f}}$ between successive states at $h_{23}=0$, $1$, and $2$ for $h_{01}=h_{12}=1$.
(e) Densities of four states as a function of $h_{20}$ at $h_{01}=h_{12}=h_{23}=h_{30}=1$.
(f)  Flow rate $q_{\mathrm {f}}$ between successive states at $h_{01}=h_{12}=h_{23}=h_{30}=h$.
The solid and dashed lines represent the data for $h=1$ and $1.5$, respectively.
From top to bottom, $q_{01}(=q_{12})$, $q_{20}$, and $q_{23}(=q_{30})$, where $q_{k[k+1]}= w_{k[k+1]}\rho_k - w_{[k+1]k}\rho_{[k+1]}$.
}
\label{fig:fix}
\end{figure*}

Before simulating the lattice Potts models,
we investigate the flip dynamics in homogeneous mixed states.
It corresponds to a lattice model with no interactions between neighboring sites 
or with strong mixing, like in a continuous flow stirred tank reactor (CSTR) in chemical reaction.\cite{hohm96}
Each site flips independently, and hence,
the density $\rho_i$ of the $s=i$ state is followed by the equation,
\begin{equation}\label{eq:st4t}
\frac{d\rho_i}{dt} = \sum_{j\ne i} w_{ji}\rho_j - w_{ij}\rho_i,
\end{equation}
where $w_{ij}$ is the flip rate from $s=i$  to $s=j$
and $\sum_i \rho_i = 1$.
We consider two types of flip routes as shown in Fig.~\ref{fig:fix}a and b.
In the first case (Fig.~\ref{fig:fix}a), each state can flip only to successive states ($s=k$ to $s=[k-1]$ or $s=[k+1]$),
where $[k']= k' \mathrm{\ mod\ } 4$, i.e., $w_{02}=w_{20}=w_{13}=w_{31}=0$.
We consider the condition of $w_{k[k+1]} \geq w_{[k+1]k}>0$ for $k\in [0,3]$, and therefore,
each site exhibits a cyclic change in the four states on average as $s=0 \to 1 \to 2\to 3 \to 0$.
In the second case (Fig.~\ref{fig:fix}b), diagonal flips between $s=0$  and $s=2$ are additionally allowed
($w_{20} \geq w_{02}>0$ and $w_{13}=w_{31}=0$).
Each site can exhibit the cyclic change in the three states ($s=0 \to 1 \to 2\to 0$) 
as well as that in the four states.
Therefore, we refer to them as four-state cyclic condition
and three- and four-state cyclic condition, respectively.
Chemical reaction on a substrate is an example of possible realizations of 
these four states.
$s=0$ is the unoccupied site, a reactant molecule binds on the site at $s=1$,
and the bound molecule changes its structure by sequential reactions as $s=1\to 2 \to 3$ (see Fig.~S1).
The final product unbinds from the substrate in the four cycle ($s=3\to 0$)
and the intermediate product unbinds in the three cycle ($s=2\to 0$).

We use the Metropolis rate $w_{ij}=\min[1,\exp(h_{ij})]$
and $h_{ji}=-h_{ij}$ for available flips to ensure
that the flips between two states obey the \textit{local} detailed balance condition.
The thermal energy $k_{\mathrm{B}}T$ is normalized to unity and $h_{k[k+1]} \ge 0$.
The steady state is obtained by solving the simultaneous equations of $d\rho_i/dt =0$.
Under the four-state cyclic condition,
the steady state is given by
\begin{eqnarray}\label{eq:st4}
&&\frac{\rho_k}{\rho_{[k-1]}} = e^{h_{[k-1]k}} \times \\ \nonumber  
&&\frac{1 + e^{-h_{k[k+1]}}+ e^{-h_{k[k+1]}-h_{[k+1][k+2]}}  + e^{h_{[k-1]k}-h_{\mathrm{cyc}}} }{1 + e^{-h_{k[k+1]}}+ e^{-h_{k[k+1]}-h_{[k+1][k+2]}}  + e^{h_{[k-1]k}}\hspace{8mm}},
\end{eqnarray}
where $h_{\mathrm{cyc}}=h_{01}+h_{12}+h_{23}+h_{30}$.
This is similar to that of the three-state cyclic condition reported in Ref.~\citenum{nogu24b}.
For $h_{\mathrm{cyc}}=0$, Eq.~(\ref{eq:st4}) provides the detailed balance relation $\rho_k/\rho_{[k-1]} = e^{h_{[k-1]k}}$ in thermal equilibrium.
However, for $h_{\mathrm{cyc}}\ne 0$, the density ratio deviates from it as shown in Fig.~\ref{fig:fix}c. 
Forward flips occur more frequently than the backward flips and the flow rate $q_{k[k+1]}= w_{k[k+1]}\rho_k - w_{[k+1]k}\rho_{[k+1]}$ increases with increasing $h_{30}$, and $q_{\mathrm {f}}=q_{01}=q_{12}=q_{23}=q_{30}$ in the steady state (see Fig.~\ref{fig:fix}d).

Under the three- and four-state cyclic condition,
the cyclic flips in the three states ($s=0 \to 1 \to 2 \to 0$)
increase with increasing $h_{20}$ (see Fig.~\ref{fig:fix}e and f).
The total flow $q_{12}= q_{23}+q_{20}$ also increases,
while the flow $q_{23}=q_{30}$ through the four states decreases.

\begin{figure*}[t]
\includegraphics[]{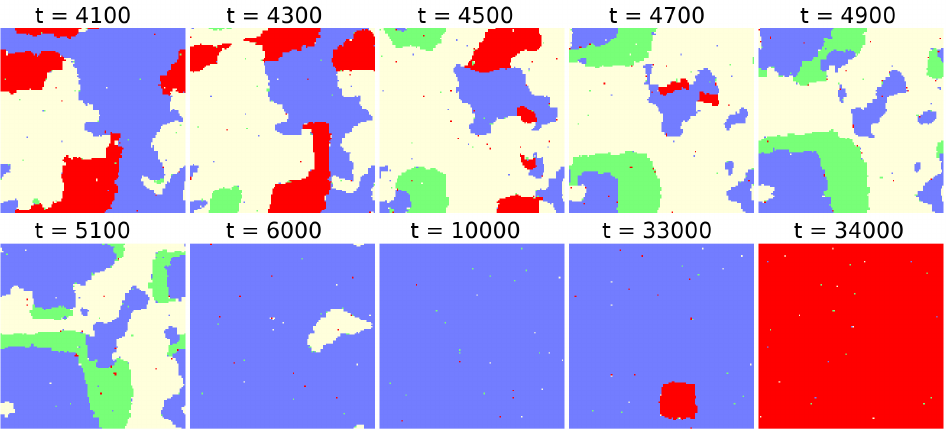}
\caption{
Sequential snapshots in the symmetrically cyclic condition at $h_{01}=h_{12}=h_{23}=h_{30}=h=0.9$ and $N=128^2$.
The light yellow, light green, medium blue, and red sites (light to dark in grayscale) 
represent $s=0$, $1$, $2$, and $3$, respectively.
Four states spatially coexist at $t\leq 4700$. At $t>10000$, 
one state is dominantly occupied with the entire system, and the dominant state cyclically changes via nucleation and growth.
}
\label{fig:snapm09s}
\end{figure*}

\begin{figure*}[t]
\includegraphics[]{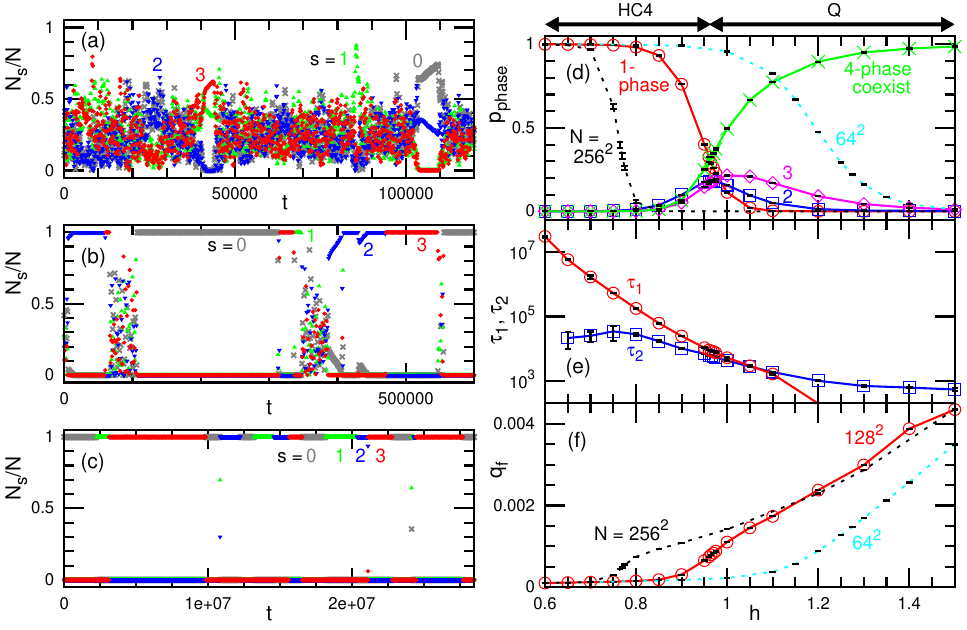}
\caption{
Dynamics in the symmetrically cyclic condition ($h_{01}=h_{12}=h_{23}=h_{30}=h$).
(a--c)  Time evolution of the fraction of sites in each state
for (a) $h=1.2$, (b) $h=0.9$, and (c) $h=0.7$ at $N=128^2$.
The data in (b) correspond to the snapshots in Fig.~\ref{fig:snapm09s}.
(d--f) Dependence on $h$.
The solid lines represent the data at $N=128^2$.
(d) Probabilities of the one-phase and multi-phase coexistence states.
The black and light-blue dashed lines represent the one-phase probabilities at $N=256^2$ and $N=64^2$, respectively.
The upper arrows indicate the region of the HC4 and Q modes at $N=128^2$.
(e) Mean lifetime of one phase ($\tau_1$) and coexistence of diagonal two phases ($\tau_2$).
(f) Flow rate $q_{\mathrm{f}}$ between successive states.
The black and light-blue dashed lines represent the data at $N=256^2$ and $N=64^2$, respectively.
}
\label{fig:m1s}
\end{figure*}

\subsection*{Symmetric Four-State Cycling}

We consider the Potts model of a two-dimensional (2D) square lattice consisting of $N$ sites.
The nearest neighboring sites have contact energy $-J$ if they are in the same state and $0$ 
if they are in different states.
We use $J=2$ to induce phase separation between different states.
The dynamics of the Potts model under the four-state cyclic condition (Fig.~\ref{fig:fix}a) are presented below.

\begin{figure*}[ht!]
\includegraphics{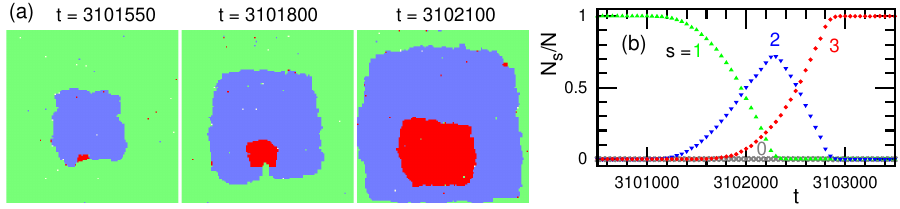}
\caption{
Transition from the $s=1$ dominant phase to the $s=3$ dominant phase in the HC4 mode 
under the cyclically  symmetric condition at $h_{01}=h_{12}=h_{23}=h_{30}=h=0.7$ and $N=128^2$ (corresponding to the data shown in Fig.~\ref{fig:m1s}c).
(a) Sequential snapshots. 
(b) Time evolution of the fraction of sites in each state.
The nucleation and growth of an $s=2$ domain occur, and meanwhile,
an $s=3$ domain appears at the domain boundary and grows in the $s=2$ domain.
}
\label{fig:t07dn}
\end{figure*}

We first consider the symmetric condition, $h_{01}=h_{12}=h_{23}=h_{30}=h$ (Figs.~\ref{fig:snapm09s}--\ref{fig:t07dn}).
At low $h$, the lattice is dominantly occupied by one of the states most of the time,
and the dominant phases cyclically change as $s=0 \to 1 \to 2 \to 3 \to 0$. 
Conversely, at high $h$, the domains of the four states spatially coexist.
We call these two modes HC4 (homogeneous cycling of the four phases) and Q (quad-phase coexistence), respectively.
At medium $h$, these two modes temporally coexist, as shown in Figs.~\ref{fig:snapm09s} and \ref{fig:m1s}b and Movie S1.
These behaviors are similar to those observed in the three-state Potts model~\cite{nogu24a};
however, there are clear differences:
i) The spatial coexistence of the four states does not form stable spiral waves unlike that in the three-state Potts model;
three domain boundaries can stably meet at one point in the 2D space, whereas four boundaries do not. 
Thus, the domain boundary of successive states ($s=k$ and $[k+1]$) moves in the direction from the $s=[k+1]$ to $s=k$ domains,
but the resultant waves do not have a center.
Since the diagonal states ($s=0$ and $2$ or $s=1$ and $3$) do not flip to each other,
the domain boundary of the diagonal states does not move ballistically but exhibits slow diffusion through the nucleation of the other states at the boundary (the diagonal boundaries look stopping in Movie S1).
ii) Spatial coexistence of the diagonal-state domains.
Once the domains of $s=k$ state stochastically disappear ($s=3$ at $t=4900$ in Fig.~\ref{fig:snapm09s});
the $s=[k-1]$ state spreads to the entire space. During the spreading, the domain of the diagonal state $s=[k+1]$ has a long lifetime (see the snapshot at $t=6000$ in Fig.~\ref{fig:snapm09s} and the state fraction at $t \simeq 500~000$ in Fig.~\ref{fig:m1s}b).
 The circular domains of the diagonal state shrink slowly, reducing the boundary energy. 
Note that a quasi-straight domain boundary connected to itself through a periodic boundary has a longer lifetime, 
since it has no preferred direction.
When the domains of the other states grow before spreading out, the system goes back to the Q mode (see the state fraction at $t \simeq 110~000$ in Fig.~\ref{fig:m1s}a).
iii) In the HC4 mode, the sequence of the dominant phases is occasionally skipped via the nucleation of the next state during the domain growth (see Figs.~\ref{fig:m1s}c and \ref{fig:t07dn}). For example, the formation of an $s=3$ domain at the boundary of a circular $s=2$ domain results in the $s=3$ dominant phase (see Fig.~\ref{fig:t07dn}a and Movie S2). The target pattern of $s=3$ and $s=2$ domains is formed in the $s=1$ phase, since the growing $s=2$ domain surrounds the unmoving domain boundary between $s=1$ and $s=3$. In contrast, the contact of the three domain boundaries results in the formation of spiral waves in the three-state Potts model.
Thus, the existence of this inactive domain boundary is the origin of the dynamics different from those in the three-state Potts model.

To quantitatively clarify the mode transitions, we calculate the time fraction of each phase (see Fig.~\ref{fig:m1s}d). 
The lattice is considered to be covered by one phase at $N_s/N>0.98$ for $s \in [0,3]$,
and $n$ phases spatially coexist when $n$ states satisfy $N_s/N>0.05$.
As $h$ increases, the time fraction of one-phase existence decreases and that of the four-phase coexistence increases; they correspond to the fraction of the HC4 and Q modes, respectively.
Hence, we consider the dominant mode to be the HC4 (Q) mode when the one-phase fraction is larger (smaller) than the four-phase fraction (see the solid lines and upper arrows in Fig.~\ref{fig:m1s}d).
As the system size $N$ increases, the transition between the HC4 and Q modes occurs at lower $h$ (see the dashed lines in Fig.~\ref{fig:m1s}d and the solid line in Fig.~S2).
This is because the nucleation occurs more frequently ($\propto N$) and the complete disappearance of one of the states occurs less  frequently in larger systems.

\begin{figure*}[t]
\includegraphics[]{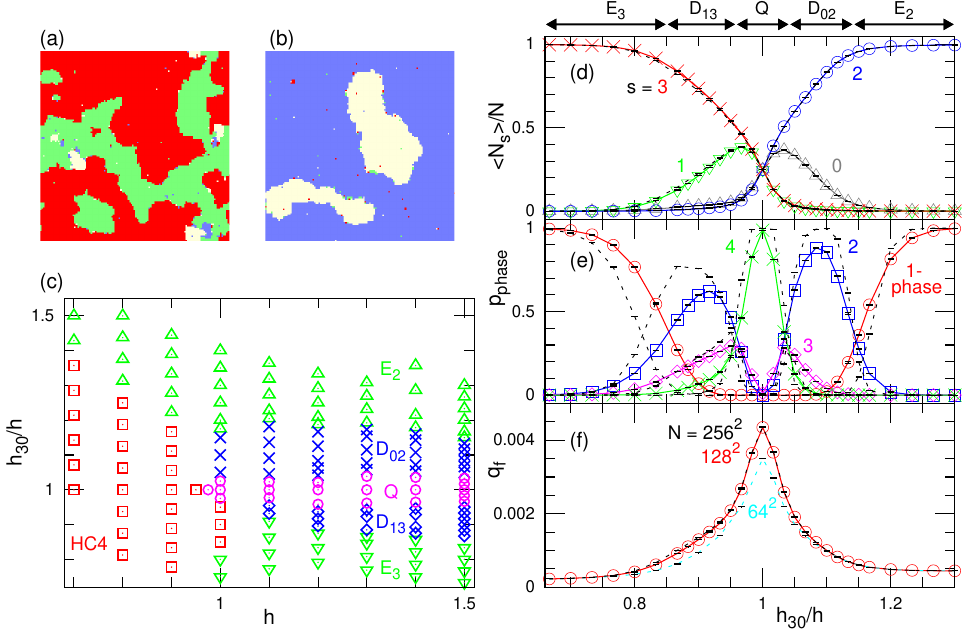}
\caption{
Dynamics in the four-state cyclic condition with $h_{01}=h_{12}=h_{23}=h$.
(a--b) Snapshots at $h=1.5$ and $N=128^2$.
(a) D$_{13}$ mode at $h_{30}/h=0.9$.
(b) D$_{02}$ mode at $h_{30}/h=1.1$.
(c) Dynamic phase diagram at $N=128^2$.
Red squares: HC4. Magenta circles: Q. Green upward- and downward-pointing triangles: E$_2$ and E$_3$.
Blue crosses and diamonds: D$_{02}$ and D$_{13}$.
(d--f) Dependence on $h_{30}$ at $h=1.5$.
The solid and black dashed lines represent the data at $N=128^2$ and $N=256^2$, respectively.
(d) Fraction $N_s/N$ of sites in each state.
The upper arrows indicate the region of the modes at $N=128^2$.
(e) Probabilities of one-phase state and multi-phase coexistence states.
(f) Flow rate $q_{\mathrm{f}}$ between successive states.
The light-blue dashed line represents the data at $N=64^2$.
The black dashed lines ($N=256^2$) in (d) and (f) overlap with the solid lines.
}
\label{fig:m1b}
\end{figure*}

Since the nucleation also occurs more frequently at higher $h$, the mean lifetime $\tau_1$ of the one-phase decreases exponentially with increasing $h$ (see Fig.~\ref{fig:m1s}e).
The mean lifetime $\tau_2$ of the two-phase coexistence decreases more slowly and becomes longer than $\tau_1$ at high $h$ in the Q mode (compare two lines in Fig.~\ref{fig:m1s}e); hence, a two-phase coexistence temporally appears in the Q mode as shown in Fig.~\ref{fig:m1s}a.
The cycling of states is considerably faster in the Q mode than that in the HC4 mode,
since the domain boundaries, in which the flips occur without an energy penalty, constantly exist in the Q mode (see Fig.~\ref{fig:m1s}f).

\subsection*{Asymmetric Four-State Cycling}

To examine the effects of asymmetric flip energies, we varied $h_{30}$ while keeping the other parameters constant at $h_{01}=h_{12}=h_{23}=h$ (see Fig.~\ref{fig:m1b}).
At high or low $h_{30}/h$, one of the states  dominates the entire system and does not cyclically change to other phases, like the dominant phases in thermal equilibrium.
We call this steady state E$_k$ mode, where $s=k$ is the dominant state.
We distinguish the HC4 and E$_k$ modes using the probability distribution of $N_s/N$ (see Fig.~S3).
In the HC4 mode, all of states have a peak at $N_s/N \simeq 1$.
When the peak of one state at $N_s/N \simeq 1$ does not exist or is too low (less than ten times of the local minimum close to $N_s/N = 1$), we consider it as the E$_k$ mode, where the $s=k$ state has the highest peak.
Further, the coexistence of the diagonal two phases appears between the Q and E$_k$ modes,
which we call D$_{k[k+2]}$ mode ($k=0$ or $1$, see Fig.~\ref{fig:m1b}a and b and Fig.~S3c).
We consider that the system is in the D$_{k[k+2]}$ mode when the time fraction of the two-phase coexistence of the $s=k$ and $s=[k+2]$ states is larger than those of the one-phase and four-phase (see Fig.~\ref{fig:m1b}e).
The D$_{k[k+2]}$ modes do not appear in the three-state Potts model,
whereas the E$_k$ modes do.

\begin{figure*}[t]
\includegraphics[]{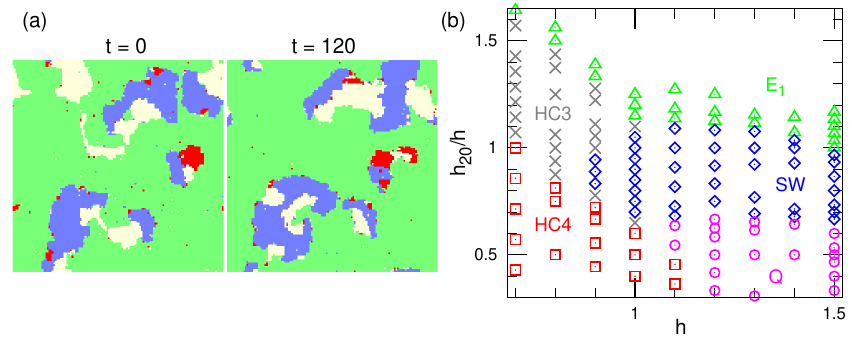}
\caption{
Dynamics in the three- and four-state cyclic condition with $h_{01}=h_{12}=h_{23}=h_{30}=h$ at $N=128^2$.
(a) Sequential snapshots at $h=1.5$ and $h_{20}/h=0.87$.
(b) Dynamic phase diagram.
Red squares: HC4. Magenta circles: Q. Green triangles: E$_1$.
Gray crosses: HC3. 
Blue diamonds: SW.
}
\label{fig:pdws}
\end{figure*}

At low $h$, the HC4 mode changes into the E$_2$ mode with increasing $h_{30}$ (see Fig.~\ref{fig:m1b}c).
This is different from that in the homogeneously mixed system, in which $N_0$ increases and $N_3$ decreases (see Fig.~\ref{fig:fix}c).
This dependency is similar to that of the three-state Potts model~\cite{nogu24b} (see the next subsection),  
but the mechanism is different.
The lifetime $\tau_1$ of the $s=3$ dominant phase decreases with increasing $h_{30}$,
whereas those of the other phases are almost constant at $h_{30}/h<1.4$ (see Fig.~S4b).
The transition from the $s=k$ to $s=[k+1]$ dominant phases is caused by nucleation and growth.
At high $h_{30}$, an $s=3$ domain in the $s=2$ phase changes into an $s=0$ domain through the nucleation in the $s=3$ domain (see Fig.~S4c). When the resultant $s=0$ domain in the $s=2$ phase is circular, it slowly shrinks, and the system returns back to the $s=2$ phase; therefore, the cycling fails.
This returning dynamics occurs more frequently at higher $h_{30}$.
Thus, the fraction of the $s=2$ phase increases, and the E$_2$ mode is eventually formed (see Fig.~S4a).

\begin{figure*}[t]
\includegraphics[]{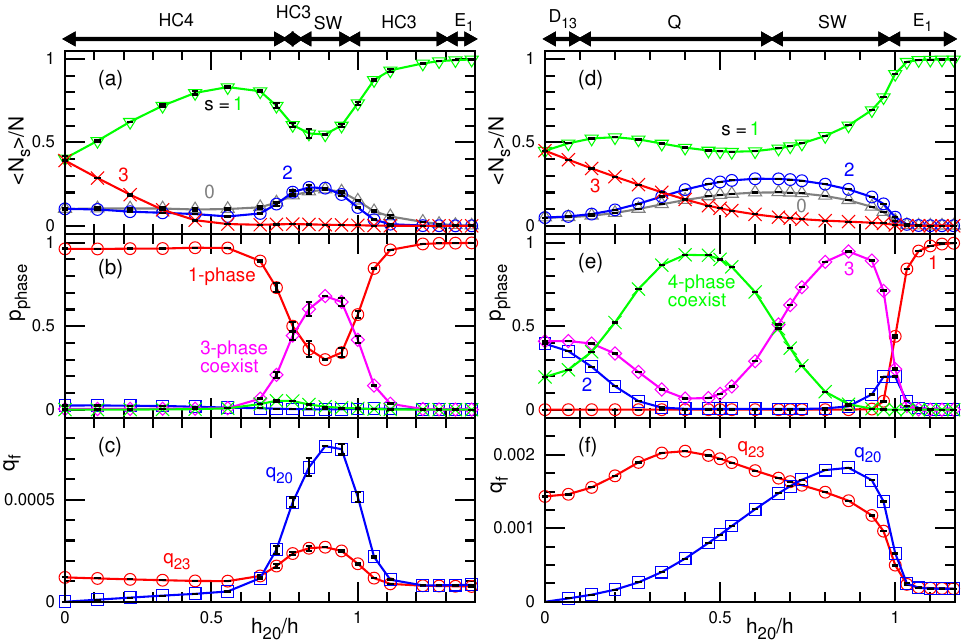}
\caption{
Dependence on $h_{20}$ in the three- and four-state cyclic condition with $h_{01}=h_{12}=h_{23}=h_{30}=h$ at $N=128^2$.
(a--c) $h=0.9$. (d--f)  $h=1.5$.
(a), (d) Fraction of sites in each state.
(b), (e) Probabilities of one-phase state and multi-phase coexistence states.
(c), (f) Flow rate $q_{\mathrm{f}}$ between successive states.
The red circles and blue squares represent $q_{23}$ and  $q_{20}$, respectively.
The upper arrows indicate the region of the modes.
}
\label{fig:mws}
\end{figure*}

At high $h$, the mode sequentially changes from E$_3$, to D$_{13}$, Q, D$_{02}$, and E$_2$ with increasing $h_{30}$ (see Fig.~\ref{fig:m1b}c--e).
With increasing system size,  the region of each mode remains almost unchanged but the mode boundaries become sharper (narrower regions of mode coexistence), as shown in Fig.~\ref{fig:m1b}e. 
The flow rate $q_{\mathrm{f}}$ is maximized in the Q mode at $h_{30}/h=1$ (see Fig.~\ref{fig:m1b}f).
In the D$_{13}$ and D$_{02}$ modes, $s=1$ and $s=0$ states form smaller domains that shrink slowly, respectively. 
New domains are formed by the nucleation and growth of the other states ($s=0$ and $2$ for D$_{13}$ and $s=1$ and $3$
for D$_{02}$), as shown in Fig.~S5 and Movie S3.

\subsection*{Three-State and Four-State Cycling}

We consider the Potts model for the three- and four-state cyclic condition (Fig.~\ref{fig:fix}b).
We fix the four-state cyclic condition as $h_{01}=h_{12}=h_{23}=h_{30}=h$ and vary $h_{20}$ of the three-state cycle.
As expected, the homogeneous-cycling mode of the three states (HC3, $s=0\to 1 \to 2 \to 0$)
and spiral-wave mode (SW) appear with increasing $h_{20}$ (see Fig.~\ref{fig:pdws}).
In the SW mode, the domains of the three states ($s=0$, $1$, and $2$) rotate around the contact point of the three domains (see Fig.~\ref{fig:pdws}a and Movie S4).
We consider the system to be in the SW mode if the time fraction of the coexistence of $s=0$, $1$, and $2$ phases is larger than those of the others (see Fig.~\ref{fig:mws}).
The flow through the three states ($q_{20}$ for $s=0\to 1 \to 2 \to 0$) is faster than that through the four states ($q_{23}$ for $s=0\to 1 \to 2 \to 3 \to 0$)  in the SW mode (see Fig.~\ref{fig:mws}c and f).
When the one-phase fraction is the largest, if the $s=0$, $s=1$, and $s=2$ states have peaks at $N_S/N\simeq 1$ but $s=3$ does not,
it is in the HC3 mode (see Fig.~S6).

At low $h$, the HC4 mode changes into the HC3 mode, and subsequently to the E$_1$ mode, with an increase in $h_{20}$ (see Figs.~\ref{fig:pdws}b and S7).
The E$_1$ mode is formed due to an increase in the lifetime of the $s=1$ dominant phase (see Fig.~S7b)~\cite{nogu24b};
let us consider a cluster of two sites of $s=2$ in the $s=1$ phase.
There are two ways to reduce the cluster size. One is the direct flip of from $s=2$ to $s=1$ with the rate $\min[1,\exp(2J-h_{12})]$, since the number of the contacts between $s=1$ and $2$ sites is reduced from six to four.
The other is the two-step flip ($s=2\to 0 \to 1$) with the rates $\min[1,\exp(-J+h_{20})]$ and $\min[1,\exp(3J+h_{01})]$
for the first (rate-limiting) and second steps, respectively.
The latter indirect flip proceeds more frequently  with increasing $h_{20}$,
and hence the $s=1$ phase becomes the dominant phase rather than the $s=0$ phase.
In contrast to the mode transition from HC4 to E$_k$, the nucleation is suppressed during the three-state cycling.

\section*{Discussion}

We have studied the dynamics of the nonequilibrium four-state Potts models.
In the four-state cyclic condition, four types of modes are observed:
cyclic changes in the homogeneous phases (HC4), the spatial coexistence of the four phases (Q),
the spatial coexistence of the diagonal two phases (D$_{k[k+2]}$ for $k=0$ or $1$), and single homogeneous phases (E$_k$ for $k \in [0,3]$).
When the four states are cyclically symmetric, the HC4 mode changes into the Q mode via the temporal coexistence of the two modes with increasing flipping energies.
As the asymmetry of the flipping energies increases, the Q mode changes into the D$_{k[k+2]}$ modes and subsequently, the E$_k$ or E$_{[k+2]}$ modes.
When both four-state cycling and three-state cycling are allowed, the two modes of the three-state cyclic Potts model, homogeneous cycling of the three states (HC3) and spiral wave (SW) in the domains of the three states also appear. Mode changes from the four-state cycling to three-state cycling modes occur by varying the flipping energies.

Compared with three-state cycling,
the significant characteristics of four-state cycling is 
that there are no direct flips between the diagonal states $s=k$ and $s=[k+2]$.
Domains in the diagonal-state phase have a long lifetime, 
and the circular domains slowly shrink to reduce the domain boundary length.
This spatial pattern steadily exists in the D$_{k[k+2]}$ modes.
In the HC4 mode, 
these long-lived diagonal domains can cause the dominant phase to skip
and return to the previous dominant phase.

Here, we used the square lattice and a fixed contact energy.
Since homogeneous-cycling and spiral-wave modes can occur in an off-lattice three-state cyclic Potts model for undulating membranes\cite{nogu24d},
lattice structures are not essential to generate these dynamic modes.
However, the phase diagram can be modified by using different lattices or different values of the contact energy.

In the HC modes of both three and four cycles (HC3 and HC4),
an increase in the flipping energy $h_{k[k+1]}$ stabilizes the $s=[k-1]$ dominant phase.
In the HC3 mode, the nucleation of $s=k$ domains in the $s=[k-1]$ phase is suppressed by the two cyclic forward flips.
In the HC4 mode, the nucleation of $s=k$ states forms $s=[k+1]$ domains, and subsequently, they shrink and return into the $s=[k-1]$ phase, via slow backward flips.
In both cases, the $s=k$ domain formation fails, although the pathways are different.
Therefore, the stabilization of a dominant phase by the suppression of successive domain formation is likely a general mechanism in homogeneous cycling in nonequilibrium
and is expected to occur in the single-loop cycling of more than four states.
When multiple loops exist (e.g., multiple three cycles in five states), more complicated dynamics may occur. It is an open problem for further studies.

\section*{Methods}

We use a 2D square lattice with a side length of $\sqrt{N}$.
The site states are updated using a MC method.
A randomly selected site is flipped to other states.
For the four-state cycling (Fig.~\ref{fig:fix}a), one of the successive two states ($s=[k-1]$ or $[k+1]$ for $s=k$) is taken with $1/2$ probability.
For the four-state and three-state cycling (Fig.~\ref{fig:fix}b), one of the other states is taken with $1/3$ probability (no flips with $1/3$ probability for $s=1$ and $3$).
The new state is accepted with the Metropolis probability
\begin{equation}\label{eq:mc}
p_{s_is'_i}=\min\left(1,e^{-\Delta H_{s_is'_i}}\right),
\end{equation}
where $\Delta H_{s_is'_i}=H'_{\mathrm {int}}-H_{\mathrm{int}}-h_{s_is'_i}$ is the energy variation in the change from the old state to the new one,
and the interaction energy $H_{\mathrm{int}}= - J\sum_{\langle ij\rangle} \delta_{s_i,s_j}$.
This procedure is performed $N$ times per MC step (time unit).

The lifetime $\tau_1$ of the $s=k$ phase is calculated as the mean period to stay at $N_k/N>0.98$.
The lifetime $\tau_2$ of the two-phase coexistence is calculated as follows.
We consider that the D$_{02}$ phase is formed when $N_0/N>0.1$, $N_2/N>0.1$, $N_1/N<0.05$, and $N_3/N<0.05$ are satisfied,
and it is dissolved at $N_0/N<0.05$, $N_2/N<0.05$, $N_1/N>0.1$, or $N_3/N>0.1$.
Similarly for the D$_{13}$ phase.
The time averages are taken for $10^8$--$10^9$ steps, and
the statistical errors are calculated from three or more independent runs.

\begin{acknowledgments}
This work was supported by JSPS KAKENHI Grant Number JP24K06973.
\end{acknowledgments}


%

\end{document}